\documentclass[a4paper]{article}
\usepackage{epsfig}
\usepackage{amsmath}
\usepackage{amsfonts}
\usepackage{amssymb}
\usepackage{graphicx}
\usepackage{bm}
\usepackage{color}
\usepackage{latexsym}

\begin{document}

\title{Emergence of resonances in neural systems: the interplay between threshold adaptation and short-term synaptic plasticity}
\author{Jorge F. Mejias and Joaqu\'{\i}n J. Torres\\
Department of Electromagnetism and Matter Physics and \\
Institute  {\em Carlos I } for Theoretical and Computational Physics,\\
University of Granada, E-18071 Granada, Spain}
\maketitle

\begin{abstract}
In this work we study the detection of weak stimuli by spiking neurons in the presence of certain level of noisy background neural activity. Our study has focused in the realistic assumption that the synapses in the network present activity-dependent processes, such as short-term synaptic depression and facilitation. Employing mean-field techniques as well as numerical simulations, we found that there are {\em two} possible noise levels which optimize signal transmission. This new finding is in contrast with the classical theory of stochastic resonance which is able to predict only one optimal level of noise. We found that the complex interplay between the nonlinear dynamics of the neuron threshold and the activity-dependent synaptic mechanisms is responsible for this new phenomenology. Our results are confirmed by employing a more realistic FitzHugh-Nagumo neuron model, which displays threshold variability, as well as by considering more realistic synaptic models. We support our findings with recent experimental data of stochastic resonance in the human tactile blink reflex.
\end{abstract}


\section{Introduction}

It is known that a certain level of noise can enhance the detection of weak input signals for some nonlinear systems. This phenomenom, known as stochastic resonance (SR), is characterized by the presence of a resonance peak, or a bell-shaped dependence, in the information transfer measurement as a function of the noise intensity. Stochastic resonance has been measured in a wide variety of systems, including bidirectional ring lasers \cite{wiesenfeld88}, electronic circuits~\cite{fauve83}, and also in biological systems, such as crayfish mechanoreceptor~\cite{wiesenfeld94}, or voltage-dependent ion channels~\cite{bezrukov95}. In the brain, it has been found in different types of sensory neurons~\cite{MossPRL91,MossPRL00}, in the hippocampus~\cite{stacey00}, in the brain stem~\cite{hanggi08}, and in some cortical areas~\cite{chialvo93,destexhe00,manjarrez02,rudolph03}. Although this resonant behaviour in neural systems has been extensively studied, the role that realistic neural features (such as nonlinearities in neuron excitability or accurate synaptic dynamics) could have in signal detection in noisy environments is still unclear. It is known, for instance, that actual synapses present activity dependent short-term mechanisms that may strongly modify the postsynaptic neural response in a nontrivial way. Short-term depression (STD) and facilitation (STF) are two particularly relevant mechanisms that are usually present in activity dependent or {\em dynamic} synapses~\cite{AVSN97,tsodyksPNAS}. The former of these mechanisms considers that the amount of neurotransmitter ready to be released ---due to the arrival of an action potential (AP)--- is limited. Thus, the synapse needs some time to recover these resources in order to transmit the next incoming AP. As a consequence, the postsynaptic current decreases for high input frequencies, producing a nonlinear effect in the postsynaptic response. On the other hand, synaptic facilitation is related with the influx of calcium ions through voltage-sensitive channels every time a presynaptic AP arrives to the synapse which increases the cytosolic calcium concentration. In addition, it is well known that cytosolic calcium ions can bind to some sensors, near the ready-releasable pool of neurotransmitter vesicles, and increase the release probability~\cite{bertramJNEURO}. As a consequence, during the arrival of consecutive presynaptic APs, a residual calcium from the first AP is added to the following influx of calcium due to a second AP, which favours the neurotransmitter vesicle depletion in the next release event. This yields to an increase (the so called synaptic facilitation) in the postsynaptic current. 

Both mechanisms, synaptic STD and STF can work together and interact in a non trivial way during synaptic transmission. For instance, it is well known that these two mechanisms play an important role in several complex phenomena in the brain, such as synchrony and selective attention~\cite{BT05}, and in the appearance of switching behaviour between different patterns of neural activity~\cite{torresNC2007}. In particular, they may be highly relevant in signal detection in noisy environments, as for instance in cortical gain control~\cite{AVSN97} or in spike coincidence detection, as recent studies suggest~\cite{mejiasCD07}. 

Another important issue to consider during detection of weak signals by actual neural systems is neuronal threshold adaptation. In primary visual cortex, for instance, neuronal adaptation seems to be responsible for {\em contrast adaptation} (via a strong hyperpolarization caused by high contrast stimuli)~\cite{wang03,ahmed97,dragoi00,greenlee88}, or for the scaling adaptation to varying stimuli in somatosensory cortex~\cite{garcia-lazaro07}. Neural adaptation can be understood as a consequence of the nonlinearities presented in the neuron membrane dynamics, a feature that has been captured (to some extent) by a large number of neuronal models~\cite{izhikevich04}, and in particular, it can be seen as neuron threshold adaptation~\cite{wang01,longtin04}. Similarly to what occurs with dynamic synapses, threshold adaptation can yield complex and new emergent cooperative phenomena when considered in large populations of neurons, as well~\cite{horn89,torresNC2007}. However, the complex interplay between dynamic synapses and threshold adaptation has caught little attention from researchers, despite the computational implications that each one can have in different neural systems when they are considered separately.

In this work, we use a phenomenological model of dynamic synapses and a standard integrate-and-fire (IF) neuron model with variable threshold to study the interaction between threshold adaptation, STD and STF in the detection of weak (subthreshold) signals under a noisy environment. More precisely, we consider a system of $N$ presynaptic neurons which transmit APs, within a Poisson distribution with mean frequency $f_n$, to a postsynaptic neuron through dynamic synapses. In these conditions, a weak and low-frequency signal is also transmitted to the postsynaptic neuron to study its response and the conditions in which SR occurs. Our results show that new phenomena can emerge as a consequence of the interplay between threshold adaptation and short-term synaptic processes. Concretely, this interplay induces the appearance of a second resonance peak at relatively high frequencies, which coexists with the standard SR peak located at low frequencies. The coexistence of these two resonance peaks allows the system to efficiently detect incoming signals for two well defined network noise levels. The precise frequency at which each one of these two resonance peaks appear is determined by the particular values of the relevant parameters involved in the dynamics of the synapses. Our results are confirmed by employing a more realistic FitzHugh-Nagumo (FHN) neural model (which possess, due to its highly nonlinear dynamics, an internal threshold variability mechanism), as well as by considering more realistic synaptic models. Finally, we have compared the results of our study with recent experimental data which seems to shows two stochastic resonance peaks in the human tactile blink reflex~\cite{hanggi08}.

\section{The model}

We consider a postsynaptic neuron which receives a slow, weak external signal. For simplicity, this signal is considered periodical. At the same time, the neuron is exposed to the uncorrelated activity of a network of $N=200$ excitatory neurons, which acts as a background noise term added to the external signal. The scheme of the system is showed in figure~1.  

To simulate the membrane potential of the postsynaptic neuron, we employ the IF neuron model which can be expressed as

\begin{equation}
\tau_m \frac{dV(t)}{dt}=-V(t)+R_{in} I(t)
\label{IFneuron}
\end{equation}
where $V(t)$ is the membrane potential, $\tau_m=10~ms$ is the membrane time constant, and the neural input or excitatory postsynaptic current (EPSC) is given by $I(t)$, which is multiplied by the input resistance $R_{in}=0.1~G \Omega$. As a consequence of the input current $I(t),$ the membrane potential $V(T)$ depolarizes, and when it reaches a certain threshold $\theta$, an AP is generated and $V(t)$ is reset to its resting value (which, for simplicity, we set at $V_r=0$). After the generation of an AP, the membrane potential remains in its resting value for a short period of time, called the absolute refractory period, which we set at $\tau_{ref}=5~ms$. 

The neural input is constituted by the sum of two terms, namely $I(t)=S(t)+I_n(t)$. The first term, $S(t)\equiv d_s \sin(2\pi f_s t)$, is the input weak signal, with $f_s=5~Hz$ and $d_s=10~pA$ being its frequency and amplitude, respectively. The second term corresponds to the total synaptic current due to the $N=200$ uncorrelated presynaptic neurons, namely $I_n(t)\equiv \sum_{i=1}^N I_i (t)$. It takes into account the noisy current introduced by the other neurons in the network, and its level is controlled by the mean firing rate of the network $f_n$. This noisy current includes an activity-dependence of the synapses as proposed in the phenomenological model presented in~\cite{tsodyksPNAS}. According to this model, the state of the synapse $i$ is governed by the system of equations

\begin{equation}
\begin{array}{lll}
\displaystyle\frac{dx_{i}(t)}{dt}&=&\displaystyle\frac{z_{i}(t)}{\tau_{rec}}-{u_i}(t)\,x_{i}(t)\,\delta(t-t_{sp})
\\\\
\displaystyle \frac{dy_{i}(t)}{dt}&=&\displaystyle-\frac{y_{i}(t)}{\tau_{in}}+{u_i}(t)\,x_{i}(t)\,\delta(t-t_{sp})
\\\\
\displaystyle\frac{dz_{i}(t)}{dt}&=&\displaystyle\frac{y_{i}(t)}{\tau_{in}}-\frac{z_{i}(t)}{\tau_{rec}},\\    
\end{array}
\label{synapse}
\end{equation}
where $x_{i}(t)$,$y_{i}(t)$,$z_{i}(t)$ are the fraction of neurotransmitter in a recovered, active and inactive state, respectively (see ~\cite{tsodyksPNAS} for further details). Here, $\tau_{in}=3~ms$ and $\tau_{rec}$ are the synapse inactivation and active neurotransmitter recovery time constants, respectively. The delta functions appearing in equation ({\ref{synapse}}) take into account that an AP arrives to the synapse at fixed time $t=t_{sp}$. The variable $u_i(t)$ is an auxiliary variable such that the product $u_i(t)x_i(t)$ represents the fraction of available neurotransmitter that is released after the arrival of a presynaptic AP at time $t$ or, from a probabilistic point of view, the neurotransmitter release probability at that time. Synaptic facilitation is introduced by considering the following dynamics for $u_i(t)$:

\begin{equation}
\frac{du_i(t)}{dt}=\frac{U_{SE}-u_i(t)}{\tau_{fac}}+U_{SE}\,[1-u_i(t)]\,\delta(t-t_{sp}).
\label{udynamics}
\end{equation}
This equation considers the influx of calcium ions into the neuron near the synapse through voltage-sensitive ion channels~\cite{bertramJNEURO}. These ions usually can bind to some acceptor which gates and facilitates the release of neurotransmitters. Pure depressing synapses correspond to ${u_i}(t)=U_{SE}$ constant (which is also obtained in the limit $\tau_{fac}\rightarrow0$), where $U_{SE}$ is the neurotransmitter release probability without the facilitation mechanism. We consider that the excitatory postsynaptic current generated by the synapse $i$ is proportional to the amount of active neurotransmitter (i.e., that which has been released into the synaptic cleft after the arrival of an AP), namely $I_i(t)=A_{SE}~y_i(t)$. 

As can be easily checked in equations (\ref{synapse}-\ref{udynamics}), in activity dependent or dynamic synapses, the degree of synaptic depression and facilitation increases with $\tau_{rec}$ and $\tau_{fac},$ respectively, and these levels are also controlled by $U_{SE}.$ On the other hand, {\em static} synapses (i.e., when synapses are not activity dependent) are obtained for $\tau_{rec},\;\tau_{fac}\rightarrow 0$. 

To complete the description of the system, we assume that the firing threshold of the postsynaptic neuron has its own dynamics given by

\begin{equation}
\tau_{\theta} \frac{d\theta(t)}{dt}=-\theta(t) +\delta+R_{in} I(t),
\label{umbral}
\end{equation}
which implies an adaptation of neuron threshold with the incoming synaptic current $I(t),$ along a characteristic time scale $\tau_{\theta}=800~ms.$ The constant input parameter $\delta=2~mV$ in equation (\ref{umbral}) ensures
that the firing threshold lies above the mean input current at stationary state, and it guarantees that the output spiking activity is driven by the current fluctuations. Such kind of dynamics has been widely used to model threshold adaptation in many neural systems~\cite{longtin04,loxley07,persi04neco,wang01}. Moreover, we assume that the signal $S(t)$ is too weak to have an appreciable effect on the dynamics of the threshold, and therefore we set $I(t)=I_n(t)$ in equation (\ref{umbral}). To ensure physiological values of the neuron threshold, we also impose a minimum value for the firing threshold of $\theta_m=7~mV$.

\section{Results}

As we have mentioned before, the phenomenon of stochastic resonance has been measured in neurons under different conditions and, in particular, in the cortex~\cite{rudolph01,rudolph03,destexhe00,manjarrez02}. Using our IF neuron model with threshold adaptation, we studied the level of background noisy activity received by a postsynaptic neuron which improves its ability to detect an incoming weak signal.
This signal is considered weak in the sense that, if the level of noise is zero or sufficiently low, the neuron does not generate APs strongly correlated with the signal. In order to quantify the level of coherence between the input signal $S(t)$ and the response of the postsynaptic 
neuron, we can employ a cross-correlation function as defined in~\cite{collins95}, that is,
\begin{equation}
C_0 \equiv \left\langle S(t) R(t) \right\rangle = \frac{1}{T} \int_0^{T} S(t) R(t) dt,
\label{C_0}
\end{equation}
where $T$ is the total recording time of each trial, typically much greater than the signal period $f_s^{-1}$, and $R(t)$ is the instantaneous firing rate of the postsynaptic neuron. An example of stochastic resonance in the case of a presynaptic population with static synapses is shown in figure~2. For low noise frequencies, the neuron is not able to fire, and therefore, to detect the weak signal. This is reflected in the fact that $C_0$ takes low values. However, when the noise frequency is increased, both noise and signal terms contribute to make the system follow the signal, that is, the neuron response becomes highly correlated with the stimulus.
As a consequence of this, a maximum value of $C_0$ is reached. Beyond that point, the activity of the presynaptic neurons produces a high and noisy postsynaptic response, and the resonance parameter $C_0$ decays with its characteristic shape. 

This typical resonance behaviour appears when 
synapses do not show any fast variability in their strength, or when the variation is only due to a slow learning processes, which we do not consider here. However, we must take into account that actual synapses show activity-dependent variability at short time scales, and this feature could modify the response of the postsynaptic neuron to the signal. In particular, since STD is a mechanism that usually modulates the high frequency inputs, one can wonder about its effect in the SR curve. In fact, our results show that this effect is quite notorious as can be viewed in figure~3A. The figure shows the emergence of {\em bimodal} resonances in the presence of STD. More precisely, in addition to the standard SR peak, a second resonance peak appears at high frequencies and moves towards lower frequency values as the degree of depression increases. This second peak allows the system to efficiently detect the weak input signal among a wide range of high frequencies (note the logarithmic scale on $f_n$). Therefore, this new resonance peak reflects that the neuron is able to properly detect the incoming signal for both low and high values of the mean network rates. We also observe that the 
location of this second resonance peak has a nonlinear dependence with $\tau_{rec}.$ To better visualize this effect we plot in figure~3B the behaviour of $f^*,$ defined as the noise frequency value at which the second resonance peak is located, as a function of $\tau_{rec}.$ We can observe in this figure that data from numerical simulation agrees with our mean-field prediction. In the following and unless specifically specified, we have considered a time window of $\sim 10$ seconds for the simulations of the SR curves, and we have averaged each data point over $30$ trials.

As well as STD, the facilitation mechanism is able to modulate the intensity of the postsynaptic response in a nonlinear manner for given presynaptic conditions. Following a similar reasoning to the one considered above, we expect synaptic facilitation to have and important effect in the signal detection properties of the postsynaptic neuron under noisy conditions. This effect is shown in figure~3C, where depending on the value of $\tau_{fac}$, the resonance peak located at low frequencies can be tuned among different values of $f_n.$ It is worthy to note that the appearance of the low frequency peak is not induced by the presence of depression or facilitation mechanisms in the synapse, since it also appears for static synapses (see figure~2A). Therefore, it corresponds to the standard SR phenomena observed in many excitable nonlinear systems. However, its precise location in the frequency range is influenced by STF. Concretely,
since the effect of facilitation is to potentiate the postsynaptic response, one should expect that levels of noise which are too low to cause high $C_0$ values with static synapses would, in the presence of STF, contribute to the resonance. On the other hand, the noise frequency values which were optimal to cause SR in absence of STF, becomes too high in the presence of STF and provoke a decrease in $C_0.$ Considering these two effects together, one should expect a displacement of the first resonance peak towards lower values of $f_n$ as $\tau_{fac}$ increases, which is what we observe in simulations. Since the position of the first peak is highly sensitive to the value of $\tau_{fac}$, STF could have an important role for a precise discrimination of the network noisy activity level needed for the optimal detection of weak signals. The second peak, which is mainly caused by the depression mechanism, does not change its position when $\tau_{fac}$ is varied, due to the prevalence of the STD effect over the STF at high frequencies. The dependence of the position of the low frequency peak, namely $f^+$, with the facilitation characteristic time is shown in the figure~3D.



The appearance of these bimodal resonances is not exclusively due to the dynamical characteristics of synapses. Neural adaptation, which we have included in our model via a dynamical firing threshold, is of vital importance for the emergence of bimodal resonances. To illustrate this, we have computed sr curves for different values of $\tau_{rec}$ and an IF neuron with fixed firing threshold. The result is shown in figure~4A, where we can see that STD is not able to induce a second resonance peak when neuron threshold is considered constant. Instead of this, we found that $C_0$ does not decay from its peak value to zero for high $f_n$ values, but it stabilizes at a steady value $C_0^* (\tau_{rec}).$ Such high steady value means that some level of coherence between the weak signal and the postsynaptic response is maintained for high mean rates. It is worthy to note that, for a particular value of $\tau_{rec}$ ($500~ms$ in the figure), the value of $C_0^*$ obtained is similar to its peak value, thus allowing a good detection over a wide range of background firing rate values. 

This saturation of $C_0$ for strong enough STD, which is due to the oversimplification assumed by the IF model with fixed threshold, can be easily explained as follows. Firstly, our simulations show that, in order to have large values of $C_0,$ a necessary condition is that $\overline{I}_n\approx V_{th}/R_{in}$\footnote{If $\overline{I}_n\ll V_{th}/R_{in}$ the postsynaptic neuron is not firing at all, and if $\overline{I}_n\gg V_{th}/R_{in}$ the postsynaptic neuron is firing all the time.}, with $\overline{I}_n$ being the mean noisy input current. Secondly, in the presence of STD and for high background noise rate, the mean noisy input current $\overline{I}_n$ saturates at certain value $I_{\infty} \equiv \lim_{f_n \rightarrow \infty} \overline{I}_n $ ---see expressions for the mean and peak value of the postsynaptic current in the supplementary text--- which is infinity for $\tau_{rec}=0$ and decreases as $\tau_{rec}$ increases. Moreover, for $\tau_{rec}$ sufficiently high (strong depression), the mean noisy current is near its asymptotic value $I_{\infty}, $ for a finite and  relatively low noise frequency $f_n.$ As a consequence, there is a sufficiently high value of $\tau_{rec}$ for which $\overline{I}_n\approx I_{\infty}\approx V_{th}/R_{in}.$ In this situation an optimal $C_0$ value will be maintained over a wide range of network firing rates, as the figure~4A shows. 

Since short-term synaptic mechanisms alone are not able to induce bimodal resonances in simple IF neurons with fixed threshold, as we have already seen, the origin of this two-peak resonant behaviour must emerge from the interplay between these synaptic mechanisms and neural adaptation. We can sketch a simple explanation of such cooperative effect by considering that, for an excitable system displaying SR, a resonance peak is obtained when the strength of the fluctuations is approximately equal to some potential barrier height \cite{wiesenfeld89}. That is, if we define in our system the barrier height as $\Delta \Phi \equiv \theta-R_{in} \overline{I}_n$, a resonance peak will appear each time the condition $R_{in} \sigma_n \simeq \Delta \Phi$ is satisfied. Considering a threshold dynamics such as the one defined in equation (\ref{umbral}), the barrier height can be approximated in the stationary state by a small constant ($\Delta \Phi \equiv \Delta \Phi_d \simeq \delta$) for large enough $f_n$. Since the dependence of $R_{in} \sigma_n$ with $f_n$ is non-monotonic for dynamic synapses (see the appendix for details), plotting together the expressions of $R_{in} \sigma_n$ and $\Delta \Phi_d$ as a function of $f_n$ shows two well located crossing points, as the top panels of figure~4B illustrate. Each one of these crossing points is associated then with a maximum in $C_0$ (as we have argued above), and therefore a bimodal resonance is obtained. The local minimum in $C_0$ is due to a high number of erratic firings of the postsynaptic neuron, which is caused by high values of the fluctuations (compared with the barrier height) around the point where the local minimum appears. This feature is depicted in the top-left panel of figure~4B with a double-head arrow. Without such large fluctuations, the local minimum of $C_0$ would vanish and the bimodal resonance would be lost. For the case of an IF neuron with static threshold, the barrier height $\Delta \Phi \equiv \Delta \Phi_s$ is a monotonically decreasing function of $f_n$. In these conditions, a single crossing point between $R_{in} \sigma_n$ and $\Delta \Phi_s$ is obtained\footnote{For certain sets of values of the model parameters, two crossing points between the level of fluctuations and the barrier height can be also found for a fixed neuron threshold. However, in such situations $\Delta \Phi_s$ is large and comparable with $\sigma.$ As a consequence, the local minimum of $C_0$ cannot be obtained, and the SR curve remains with the characteristic single-peak shape.}, and therefore the SR curve presents a single peak, as the bottom panels of figure~4B show.


The appearance of bimodal resonances gives a high versatility to neurons as weak signals detectors. In actual neural media, populations of neurons could take advantage of such versatility, and they could use the high heterogeneity of synaptic properties~\cite{markram06} to organize groups of neurons with non-resonance, single-resonance or two-resonance peak behaviour. A phase diagram, which locates the repertoire of different behaviours in the space of synaptic relevant parameters, is shown in figure~5A. For realistic synaptic conditions, the three types of behaviour are accessible. The region P2' corresponds initially to two resonances, but the second resonance is usually located in an extremely high network rate ($f^*>200~Hz$), which means that the second resonance does not occur in realistic conditions. If we increase $\tau_{rec}$ (for a given value of $\tau_{fac}$), the system pass from a single-peak resonance behaviour (region P2') for low $\tau_{rec}$, to the bimodal resonance phase (because increasing $\tau_{rec}$ implies lowering $f^*$). After that, the system reaches a single-peak behaviour again (due to the fusion of the two peaks of the bimodal resonance into one peak). Finally, increasing $\tau_{rec}$ even more would lead to a decrement of the detection ability of the neuron, leading to the zero-resonance phase. 

The fact that we considered a simplified system allowed us to derive a theoretical approach, which confirmed the numerical results both for STD and STF, as we have already seen. However, we should consider whether bimodal resonances appear in more realistic conditions. For instance, we assumed as a first approximation that fluctuations in threshold dynamics have not a dramatic influence in the appearance of the bimodal resonances. However since the SR phenomena depends strongly on noise properties this assumption could lead us to wrong conclusions. Therefore, to test our results in more realistic conditions, we consider that the dynamic threshold is driven by the noisy EPSC $I_n(t)$, instead of being driven only by its mean value $\overline{I}_n.$ The consequences of this modification do not have a dramatic effect on the resonant behaviour of the neuron, as can be seen in figure~5B. In the presence of threshold fluctuations, STD induces the appearance of a second resonance peak, as we previously found with a deterministic dynamic threshold. This second resonance peak appears for the same range of values of $\tau_{rec}$ and $f_n$, which implies that our results are robust with a more realistic fluctuating threshold adaptation. 


The emergence of bimodal resonances is also maintained when one considers a more realistic neuron model to simulate the response of the postsynaptic neuron. Although we have employed a dynamic threshold to include some of the nonlinear features of actual neurons into the IF neuron model, it should be convenient to test our findings by considering an intrinsic nonlinear neuron model which could present this type of adaptation without additional ingredients. A common model employed in the literature to model the nonlinear excitability properties of actual neurons is the FitzHugh-Nagumo neuron model \cite{kochbook}, which can be defined as
\begin{equation}
\begin{array}{lll}
\displaystyle \tau_m \frac{dv(t)}{dt}=\tau_m \,\epsilon \,\mbox{\large $[$} v(t)\mbox{\large 
  $($}v(t)-a \mbox{\large $)$} \mbox{\large $($} 1-v(t) \mbox{\large $)$}-w(t) \mbox{\large $]$} +S(t) +R\, I_n(t)
\\\\
\displaystyle~~~\frac{dw(t)}{dt}=b\,v(t)-c\,w(t),
\end{array}
\label{FHN}
\end{equation}
where $v(t)$ represents the postsynaptic membrane potential, $w(t)$ is a slow recovery variable related with the refractory time, and $a=0.001,~b=3.5~ms^{-1},~c=1~ms^{-1},~\epsilon=1000~ms^{-1}$ are parameters of the model. With this choice of values for the parameters, the model is set in the excitable regime, the (dimensionless) voltage $v(t)=1$ corresponds to $100 mV$ and time is given in $ms$. We also consider $R=0.1~G\Omega/mV$ and $\tau_m=10~ms$. The terms $S(t)$ and $I_n(t)$ are described as before, with $d_s=5$. We have performed numerical simulations of the system presented in figure~1, but considering now this FHN model for the postsynaptic neuron. The results are shown in figure~6A, where one can see that for large enough values of $\tau_{rec}$ a bimodal resonance also appears. The location of the second peak moves towards lower values of $f_n$ as $\tau_{rec}$ increases, as it was found with the IF model with dynamic threshold. The range of values of the noisy frequency $f_n$ at which the second peak is located is also the same as with the previous models with threshold dynamics. 

It is necessary to demonstrate here that the FHN model presents several threshold adaptation properties which are similar to those we assumed for the IF neuron model with dynamic threshold. In order to check this, we define two types of temporal stimuli that the postsynaptic neuron receives: $h_1(t)$ and $h_2(t)$. The first stimulus, $h_1(t)$, consists in a train of narrow ($\sim 2~ms$) square pulses of frequency $f_s$ (that is, the signal frequency). We impose that each one of these pulses arrives to the postsynaptic neuron every time the signal $S(t)$ reaches its maximum value, namely $d_s$. Similarly, the other type of stimulus, $h_2(t)$, consists in a train of narrow ($\sim 2~ms$) square pulses also of frequency $f_s$, each one of them arriving at the postsynaptic neuron when $S(t)=-d_s$, that is, every time the signal takes its lowest value. We also set a constant input $\mu$, in such a way that the total input to the postsynaptic neuron is given by $S(t)+\mu+h_1(t)+h_2(t)$. For a given fixed value of $\mu$, we can determine the value of the neural firing threshold by increasing the strength of the stimulus $h_1(t)$ (that is, the height of the narrow pulses) until an AP is generated as a consequence of such stimulus. This measure of the firing threshold will be denoted as $\theta_1$. Similarly, we can perform a second estimation of the neuron threshold, namely $\theta_2,$ by varying the strength of $h_2(t)$ until an AP is generated in response to this second stimulus. Both estimations of the firing threshold, as a function of the constant input $\mu$, are shown in figure~6B. The figure illustrates two major features of the excitability properties of the FHN neuron model. The first one is that, independently of the value of $\mu$, both estimations give almost identical results for the value of the neural firing threshold of the FHN neuron model. Since the only distinction between the stimuli $h_1(t)$ and $h_2(t)$ is a difference in amplitude of $2~d_s,$ which is due to the signal term, this result indicates that the weak signal does not influence the value of the firing threshold (independently of the value of the constant input $\mu$). This confirms the assumption we made for the IF model in equation (\ref{umbral}). The second major feature illustrated by the figure~6B is that the value of the firing threshold varies with $\mu$ as $\theta \simeq C+\mu$, with $C$ being a constant. This dependence coincides with the steady-state value of the firing threshold obtained from equation (\ref{umbral}) (see the appendix). Therefore, the assumptions we made on the modeling of the threshold dynamics for the IF model are appropriate as is confirmed by more realistic neuron models, such as the FHN model, which incorporates nonlinear excitability properties.


The robustness and generality of our previous results can be also tested by considering a more realistic model for the activity-dependent synaptic mechanisms. For instance, until now we have treated the synapses employing a standard deterministic model for the sake of simplicity. However, it is known that real synapses have a stochastic nature~\cite{synapticnoise} and their fluctuations can play an important role in neural computation \cite{synapticnoise,zadorJN}, and therefore they should be taken into account. In particular, since the SR curves depend strongly on the noise properties, it is important to consider the additional source of noise due to synaptic fluctuations, since this could lead to a very different emergent behaviour in the system. In order to test our results, we have simulated our system using an intrinsically stochastic model of dynamic synapses presented in~\cite{delarocha05}. This model considers that each connection between neurons has a number of functional contacts, or synaptic buttons, and this number is randomly chosen (for each particular connection) following a Gaussian distribution of mean $M$ and standard deviation $\Delta_M.$ In addition, the strength of each individual synaptic button is also randomly determined following a Gaussian distribution of mean $J$ and standard deviation $\Delta_J$. The release of a neurotransmitter vesicle from a synaptic button to the synaptic cleft, when an AP arrives at the button, is modeled as a random event. After that release, the recovering of the synaptic button is considered as a probabilistic event following a Poisson distribution with a typical time $\tau_{rec}.$ This probabilistic model gives the same mean values for the EPSC, but the fluctuations differ from the previous model (see figure~6D and the supplementary material for more details). As it is shown in figure~6C, this stochastic model induces the same phenomenology during SR experiments as those for the deterministic model described by (\ref{synapse}-\ref{udynamics}). That is, for the case of static synapses, a single resonance peak at low frequencies is obtained as usual, and when $\tau_{rec}$ is increased, a second peak appears at high frequencies with the resonance peak location moving towards low noise rates. We also tested our results by considering a conductance based description of the synaptic current, leading to the appearance of bimodal resonances as in the previous cases (data not shown).


While this bimodal resonance behaviour could be difficult to measure directly in \textit{in vivo} cellular recordings, several experimental methodologies are available to study the occurrence of this phenomenology in actual systems. For instance, recent experimental studies~\cite{hanggi08,schmid06} have shown that STD has a strong impact in the detection of weak tactile signals by caudal pontine reticular nucleus (PnC) neurons. This region of the brain stem is responsible for auditory startle reflex~\cite{friauf94}, and thus an auditory noisy input may act as a noise term, enhancing the tactile blink reflex response to stimuli~\cite{hanggi08}. In addition, as well as STD, neural adaptation seems to play a role in this brain area~\cite{friauf94,pilz96}. As a consequence of these factors, PnC would be an excellent brain structure to search for the existence of bimodal resonances similar to those we have theoretically obtained. Thus, we have compared the predictions of our study with experimental data\footnote{We have employed the experimental data from Table $1$ of Ref. \cite{hanggi08}, and we obtained the values of the cross-correlation function following the same protocol that the authors employed in his work.} taken from~\cite{hanggi08}. In this work it is exposed that the ability of air-puff stimulations to an eyelid (input) to induce blinks (output) is improved by the addition of auditory white noise. To measure this improvement, a cross-correlation parameter equivalent to the one we used is employed. As one can see in figure~7, experimental data show clear evidences of two resonance peaks. These resonances are well explained with our FHN model receiving a weak signal and a noisy current modulated by STD, as indicates the good fitting between experiments and our simulation results. In order to relate the auditory noise intensity $A_n$ (measured in $dB$) with the mean firing rate $f_n$, we assumed two separate regions (for $A_n \geq 60dB$ and for $A_n<60dB$), and we also considered a linear relationship $A_n=a_1~f_n+a_2$ for each region. This distinction was done in order to separate the effect of giant PnC neurons with medium ($A_n<60dB$) and high ($A_n>60dB$) threshold~\cite{friauf94}. However, other types of dependence (such as a linear dependence for all the range of $A_n$) are also plausible and does not affect the appearance of the two resonances from the experimental data. These results suggest that PnC neurons actually employ the two-resonance phenomena to increase their ability to detect weak tactile inputs over different auditory noise levels.

\section{Conclusions}

It is widely known that noise can have relevant and positive effects in many nonlinear systems in nature. These effects include noise-induced phase transitions~\cite{toral94,kawai97}, stochastic dynamics of domain growth~\cite{sancho00}, or multiple types of stochastic resonance \cite{wiesenfeld89,wiesenfeld94,collins95}, to name a few. The particular case of stochastic resonance has been widely studied in the context of biological systems~\cite{wiesenfeld94,bezrukov95}, and the occurrence of this phenomenon in the brain is well established. More precisely, it is known that stochastic resonance mechanisms are present in many brain areas, such as the cortex~\cite{destexhe00,manjarrez02,rudolph03}, the hippocampus~\cite{stacey00}, or the brain stem~\cite{hanggi08}. Therefore, it is highly relevant to address the influence that some features present in actual neural systems could have in stochastic resonance tasks. 

Short-term synaptic mechanisms are, in this framework, a good candidate to consider. It is known, for instance, that both STD and STF play an important role in the transmission of relevant correlations between neurons~\cite{mejiasCD07}, in the temporal maintenance of information in persistent states of working memory tasks~\cite{mongillo08}, in the recall of stored memories on attractor neural networks~\cite{mejias09}, or in the switching behaviour between neural activity patterns~\cite{torresNC2007}. However, the interplay between these two mechanisms, or between them and other neural adaptation processes, has not been fully understood yet. 

In this work we have considered the role of dynamic synapses in the detection of weak signals by neurons embedded in neural networks, via a stochastic resonance formalism. To the best of our knowledge, this is the first study that shows the dramatic effect of the interplay between the dynamical nature of synapses and neuron adaptation mechanisms on the stochastic resonance properties of neurons. More precisely, we have demonstrated that this interplay originates the appearance of bimodal resonances, where the location of the resonances are related with the relevant synaptic parameters. While such bimodal resonances have been found in several natural systems~\cite{toral06,bimodal03}, their occurrence in neural media has not been reported up to date.

Recent studies~\cite{zalanyi01,hanggi08} have also suggested a relevant role of STD in neural stochastic resonance, but the emergence of bimodal resonances, which is the crucial point of our study, is missed in these works. Our findings are also supported by experimental data taken from~\cite{hanggi08}, and by other experimental works~\cite{faubert08}. Several questions should be experimentally tested, though. An interesting prediction to test is, for instance, whether STF has the effect on the first resonance peak predicted by our results. The observed dependences of the position of the peaks with the synaptic characteristic time scales could be confirmed experimentally as well. Finally, the question of how these bimodal resonances can be measured in actual cortical structures, and its effect in the collective dynamics of large cortical neural networks, constitutes an interesting issue that still remains open.

\section{Acknowledgments}
This work was supported by the \textit{MEyC--FEDER} project FIS2009-08451 and the \textit{Junta de Andaluc\'{\i}a} projects P06--FQM--01505 and P07--FQM--02725. We thank E. Lugo and J. Faubert for providing us useful experimental data, and J. Marro for useful discussions.

\section*{Appendix: Analytical derivation of the noise intensity}
In this appendix, we derive the analytical expression for the cross-correlation measure $C_0$ between the response of the postsynaptic neuron to a weak input signal in the presence of synaptic noise. First, we obtain the expressions for the noisy EPSC with dynamic synapses, for both the deterministic model and the stochastic model. After that, we obtain the expression for the mean firing rate of the IF postsynaptic neuron, needed to obtain a mean-field description of $C_0$.

We consider a population of $N$ presynaptic neurons firing uncorrelated Poisson spike trains with a certain frequency $f_n$. We assume that the synaptic current $I_i(t)$ generated by an AP arriving at time $t^*$ in a particular synapse $i$ is proportional to the fraction of active neurotransmitters in that synapse, namely, $y_i(t)$ ---cf Eq. (\ref{synapse}). In this situation the postsynaptic current at time $t=t^{*}+\tau$ is given by
\begin{equation}
I_{i}(\tau,t^*)=I_p\exp(-\tau/\tau_{in}).
\end{equation}
Considering a stimulation with a stationary Poissonian AP train, the synaptic current at $t=t^*,$ namely $I_p,$ can be substituted by an averaged stationary EPSC amplitude. One easily obtains from equations (\ref{synapse}-\ref{udynamics}) that
\begin{equation}
I_p=A_{SE} ~u_\infty x_\infty
\label{peak}
\end{equation}
where $u_\infty$ and $x_\infty$ are, respectively, the facilitation and depression variables in the stationary state,  and their expressions are given by
\begin{equation}
u_\infty=\frac{U_{SE}+U_{SE}~\tau_{fac}~f_n}{1+U_{SE}~\tau_{fac}~f_n},
\end{equation}
\begin{equation}
x_\infty=\frac{1}{1+u_\infty~\tau_{rec}~f_n}.
\end{equation}

We can compute the mean noise contribution of the current and fluctuations using the central limit theorem. The following expressions are obtained
\begin{equation}
\overline{I}_n =N f_n\tau_{in} I_p
\label{media}
\end{equation}
\begin{equation}
\sigma^2_n=\frac{1}{2} ~N f_n \tau_{in} (I_p)^2
\label{varianza}
\end{equation}
where we assumed that $\tau_{in}\ll \tau_{rec}$. Equations (\ref{media}) and (\ref{varianza}) allow to characterize the noisy input from the presynaptic neurons. The dependence of these quantities with $f_n$ is shown in figure~8. We can also consider the more realistic model presented in~\cite{delarocha05}, which takes into account the stochastic nature of synaptic release events. Following~\cite{delarocha04}, this model gives the same value for the mean current but assumes an expression for the EPSC fluctuations (for an uncorrelated noisy input) that is given by
\begin{equation}
\sigma^2_n=N M J^2 u_\infty x_\infty f_n \left[ 1+\Delta_J ^2 +\frac{u_\infty \left [M(1+\Delta_M^2)-1\right ]}{1+u_\infty \tau_{rec} f_n (1-u_\infty/2)} \right],
\end{equation}
where $M$ is the number of synaptic functional contacts, $J$ is the synaptic strength per functional contact, and $\Delta_J,~\Delta_M$ are their respective standard deviations.

With these expressions (taking the fluctuations either from the deterministic or from the stochastic model), one can obtain the mean firing rate of the postsynaptic neuron by solving the problem of calculating the escape rate of a fluctuation-driven particle with linear dynamics~\cite{tuckwell89,brunel00}. We define the quantities

\begin{equation}
y_\theta (t)= \frac{\theta(t)-R_{in} \overline{I}_n  +S(t)}{R_{in} \sigma_n}
\end{equation}
\begin{equation}
y_r (t)= \frac{V_r-R_{in} \overline{I}_n +S(t)}{R_{in} \sigma_n},
\end{equation}
and assume that the weak signal $S(t)$ evolves slowly compared with the neuron dynamics. The firing rate of the postsynaptic neuron is then given by

\begin{equation}
R(t)=\left[ \tau_{ref}+ \tau_m \int_{y_r (t)}^{y_\theta (t)} dz \sqrt{\pi} \exp (z^2) (1+{\rm erf}(z)) \right]^{-1}.
\label{rate}
\end{equation}
For the case in which we have a dynamic threshold, we can set $d \theta /dt=0$ in equation~(\ref{umbral}) to obtain the stationary condition $\theta_\infty \equiv \theta=\delta+R_{in} \overline{I}_n$. On the other hand, for the static threshold approach we set $\theta (t)=\theta_m$. Equation~(\ref{rate}), together with the expressions of the EPSC and the threshold conditions obtained above, allows to evaluate the expression~(\ref{C_0}) and obtain our mean-field approach:

\begin{equation}
C_0 (\nu)=\int^{1/f_s}_0 f_s d_s \sin(2 \pi f_s) \left[ \tau_{ref}+ \tau_m \int_{y_r (t)}^{y_\theta (t)} dk \sqrt{\pi} \exp (k^2) (1+{\rm erf}(k)) \right]^{-1} dt,
\end{equation}
where we have set $T=1/f_s$. By evaluating numerically this expression, one obtains analytical curves which can be compared with the results from numerical simulations.


\newpage
\section{Figure captions}

\textbf{ Figure 1}: Schematic plot of the system considered in our study. The postsynaptic neuron (in yellow) receives a weak input periodic signal, and is exposed to the noisy background activity of other neurons (in blue). These neurons transmit Poissonian spike trains, of frequency $f_n$, through dynamic synapses. Our aim is to determine how the synaptic properties can influence the detection of the weak signal by a postsynaptic neuron with nonlinear membrane excitability properties. 
\bigskip

\textbf{ Figure 2}:(A) Characteristic curve of SR as a function of the mean network rate $f_n$. Numerical simulations of the model (symbols) agree with our mean-field theory (solid line). (B) Several time series of the postsynaptic membrane potential which correspond to different input noise frequencies (marked with a, b, c), when the postsynaptic neuron is trying to detect a weak input signal. Here, we considered static synapses ($\tau_{rec}=\tau_{fac}=0$), $U_{SE}=0.4$, $A_{SE}=120~pA$, $f_s=3~Hz$ and a fixed threshold $\theta =10~mV$. 
\bigskip

\textbf{ Figure 3}:(A) Bimodal SR curves for several values of $\tau_{rec}$, considering $U_{SE}=0.4$ and $A_{SE}=120~pA$. The effect of STD in stochastic resonance is the appearance of a second resonance peak at certain frequency $f^*$ which decreases when $\tau_{rec}$ is increased, as it is depicted in panel (B). The inset in panel (B) also shows the fitting in a clearer logarithmic scale for the vertical axis. (C) Bimodal SR curves for several values of $\tau_{fac}$, with $U_{SE}=0.1$ and $A_{SE}=350~pA.$ The panel also illustrates a decrease of the frequency $f^+,$ at which the first resonance peak appears, as $\tau_{fac}$ is increased. This dependence is clearly depicted in (D), while the inset shows the same dependence with a logarithmic scale for the vertical axis. In all panels, data from numerical simulations are denoted with symbols, whereas lines correspond to mean-field predictions. 
\bigskip

\textbf{ Figure 4}:(A) SR curves for an IF neuron model with fixed threshold $\theta=8~mV$ receiving a weak signal and a noisy input modulated by depressing synapses, for $U_{SE}=0.5,~A_{SE}=90~pA$ and several values of $\tau_{rec}$. One can see that ignoring the threshold dynamics can lead to drastic modifications in the performance of the system (cf. figure~3A). Numerical simulations (symbols) are supported with a mean field approach (lines). (B) Schematic plot that illustrates how a resonance peak appears when the amplitude of the voltage variations induced by synaptic current fluctuations (that is, $\sigma \equiv R_{in} \sigma_n$) is comparable with the barrier height $\Delta \Phi$ (see main text). In the case of an IF neuron model with dynamic threshold and in the presence of dynamic synapses, this occurs at two frequency values separated by a frequency range where $\sigma \gg \Delta \Phi\equiv \Delta \Phi_d$ (which induces sustained spiking activity and therefore decreases the coherence $C_0$ between the two maxima). For an IF neuron model with fixed threshold, however, $\sigma$ is comparable with $\Delta \Phi_s$ only for a single frequency value which explains the emergence of a single resonance peak. 
\bigskip

\textbf{ Figure 5}:(A) Phase diagram, obtained with our mean-field approach, which shows different regimes of the behaviour of the system, for $U_{SE}=0.1$ and $A_{SE}=120~pA$. Labels P0, P1, P2 denote, respectively, regions in which zero, one, or two resonance peaks appear. The region P2' denotes values of the synaptic parameters for which a second resonance appears, but at a frequency too much high to consider in realistic conditions (that is, $f^*>1/\tau_{ref} =200~Hz$). For $\tau_{rec} \rightarrow 0$ the typical single resonance peak is recovered. (B) Numerical simulations of the SR curves for an IF neuron model with STD and a noisy threshold adaptation, for $U_{SE}=0.2,~A_{SE}=110~pA$ and different values of $\tau_{rec}$. As the figure shows, the conclusions for a deterministic dynamic threshold are maintained when we take into account threshold fluctuations. Each simulation point is averaged over $100$ trials. 
\bigskip

\textbf{ Figure 6}:(A) Numerical SR curves for a postsynaptic FHN neuron model receiving a weak signal and uncorrelated background noisy activity of frequency $f_n$, for several values of $\tau_{rec}$, $\tau_{fac}=0$, $U_{SE}=0.5$ and $A_{SE}=15~pA$. In order to estimate the firing times of the FHN model, the dynamics of the variable $v(t)$ was thresholded at $v=0.8$.(B) Estimation of the neuron firing threshold for different values of a constant input current $\mu$, and employing two different measures (see the main text for details). (C) Numerical SR curves for several $\tau_{rec}$ values and $U_{SE}=0.5$, when a more realistic stochastic model for the synapses is employed. We set the parameters of the stochastic model in $M=50$, $J=3~pA$, $\Delta_M=0.1$ and $\Delta_J=1~pA$. (D) Comparison of the standard deviation of the synaptic current for the two synaptic models employed in our study. The conditions are the same than those in panel C and $\tau_{rec}=100~ms$. The difference between these two expressions is about $60~\%$ for high frequencies, although the second resonance peak is clearly obtained with both models. 
\bigskip

\textbf{ Figure 7}: Comparison between experimental data from~\cite{hanggi08} and numerical simulations of the FHN neuron model and an stochastic dynamic synapse model with $J=3~pA,~\Delta_J=1~pA, ~M=50,~\Delta_M=0.1,~U_{SE}=0.5$ and $\tau_{rec}=500~ms$. We assumed a linear relationship between auditory noise intensity and the mean firing rate ($f_n=a_1A_n+a_2$, with $(a_1,a_2)=(6,-370)$ for $A_n \geq 60dB$ and $(a_1,a_2)=(0.1,-2.5)$ for $A_n<60~dB$), although other dependences are possible and also show good agreement between experiments and simulations with realistic parameter values. Each simulation point has been averaged over $100$ trials. The inset shows the same data in a linear scale. 
\bigskip

\textbf{ Figure 8}: Mean EPSC as a function of the mean firing rate $f_n$, with $U_{SE}=0.5$, $A_{SE}=70~pA$ and $\tau_{rec}=500~ms$. Numerical simulations (symbols) are supported by mean field results (lines). In the inset, we can see the good agreement between mean field and simulations for the EPSC fluctuations.

\newpage
\section{Figures}
\bigskip


\begin{figure}[h!]
\centerline{\psfig{file=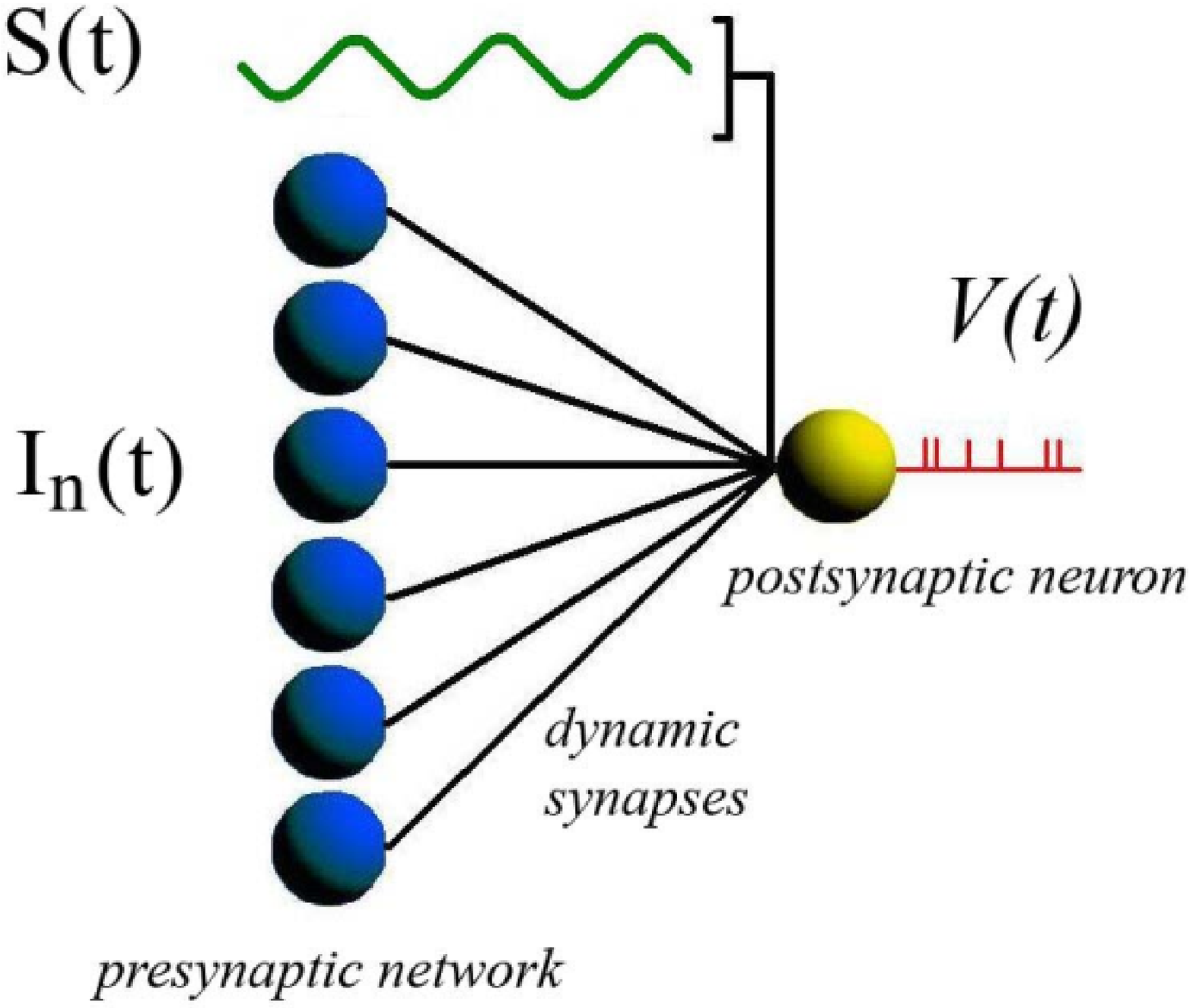,width=12cm}}
\caption{Mejias and Torres}
\label{fig1}
\end{figure}
\newpage 

\begin{figure}[t!]
\centerline{\psfig{file=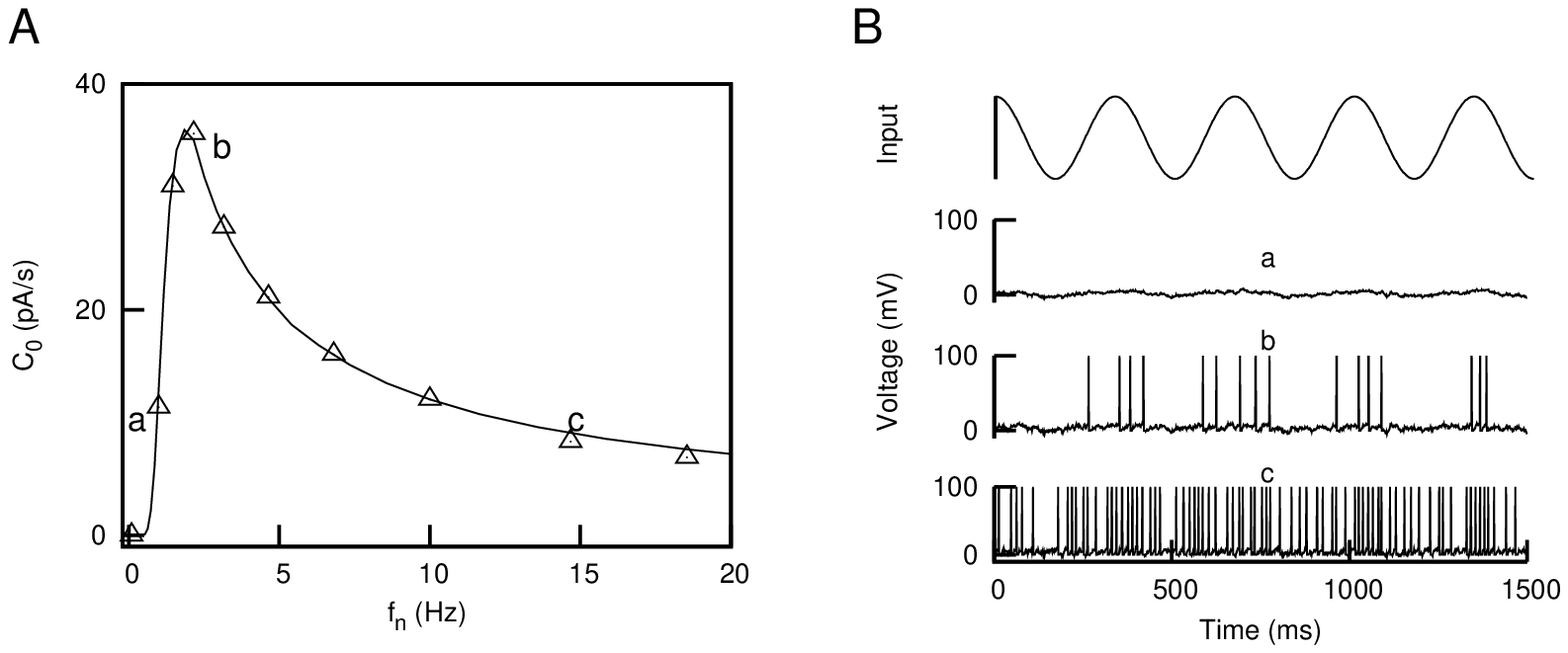,width=14cm}}
\caption{Mejias and Torres}
\label{fig2}
\end{figure}
\newpage 

\begin{figure}[t!]
\centerline{\psfig{file=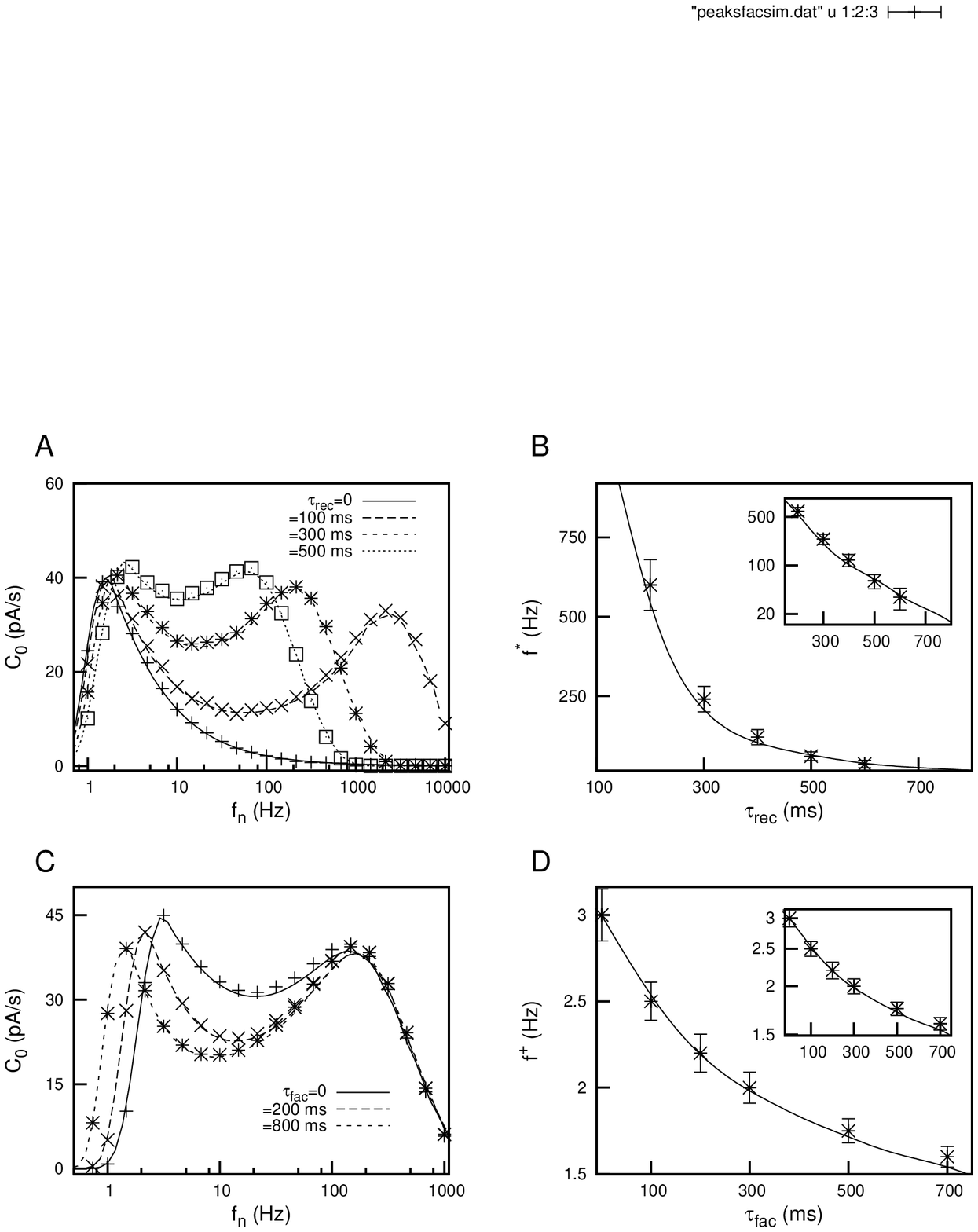,width=14cm}}
\caption{Mejias and Torres}
\label{fig3}
\end{figure}
\newpage 

\begin{figure}[t!]
\centerline{\psfig{file=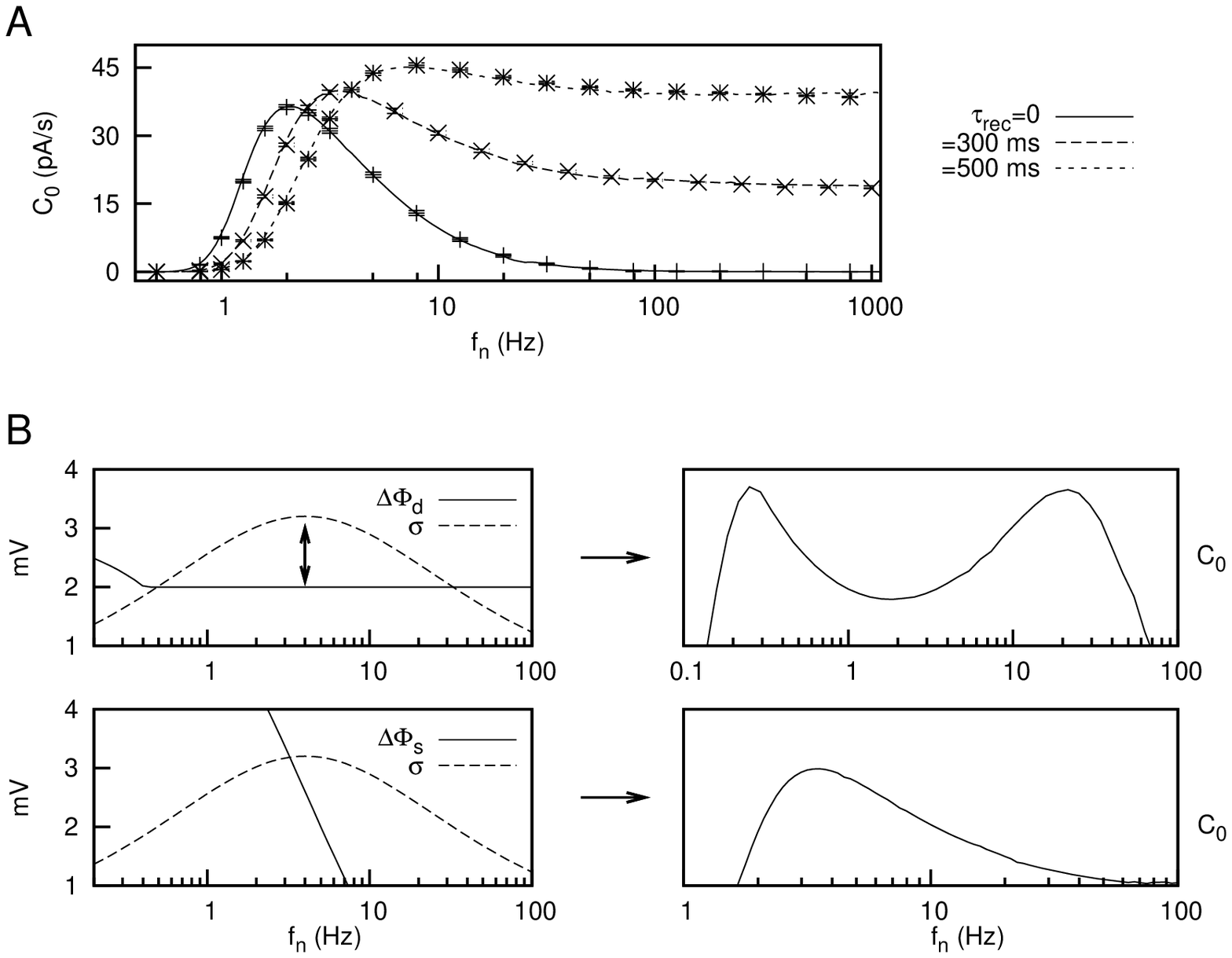,width=14cm}}
\caption{Mejias and Torres}
\label{fig4}
\end{figure}
\newpage 

\begin{figure}[t!]
\centerline{\psfig{file=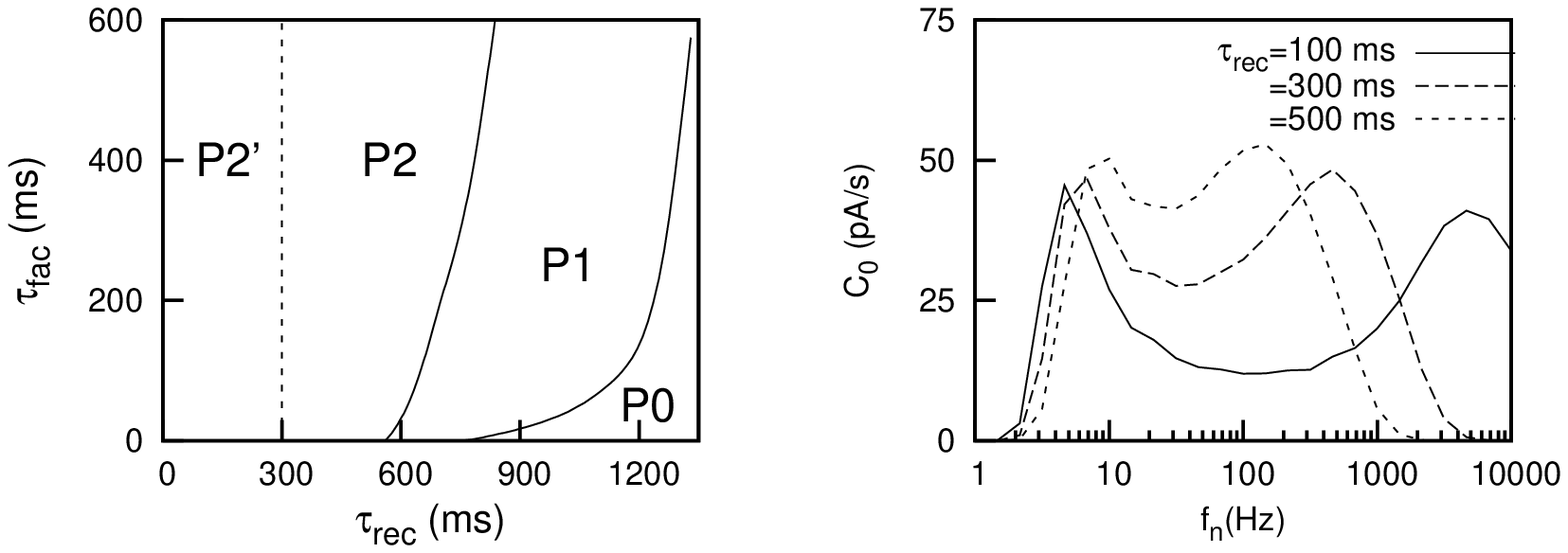,width=14cm}}
\caption{Mejias and Torres}
\label{fig5}
\end{figure}
\newpage 

\begin{figure}[t!]
\centerline{\psfig{file=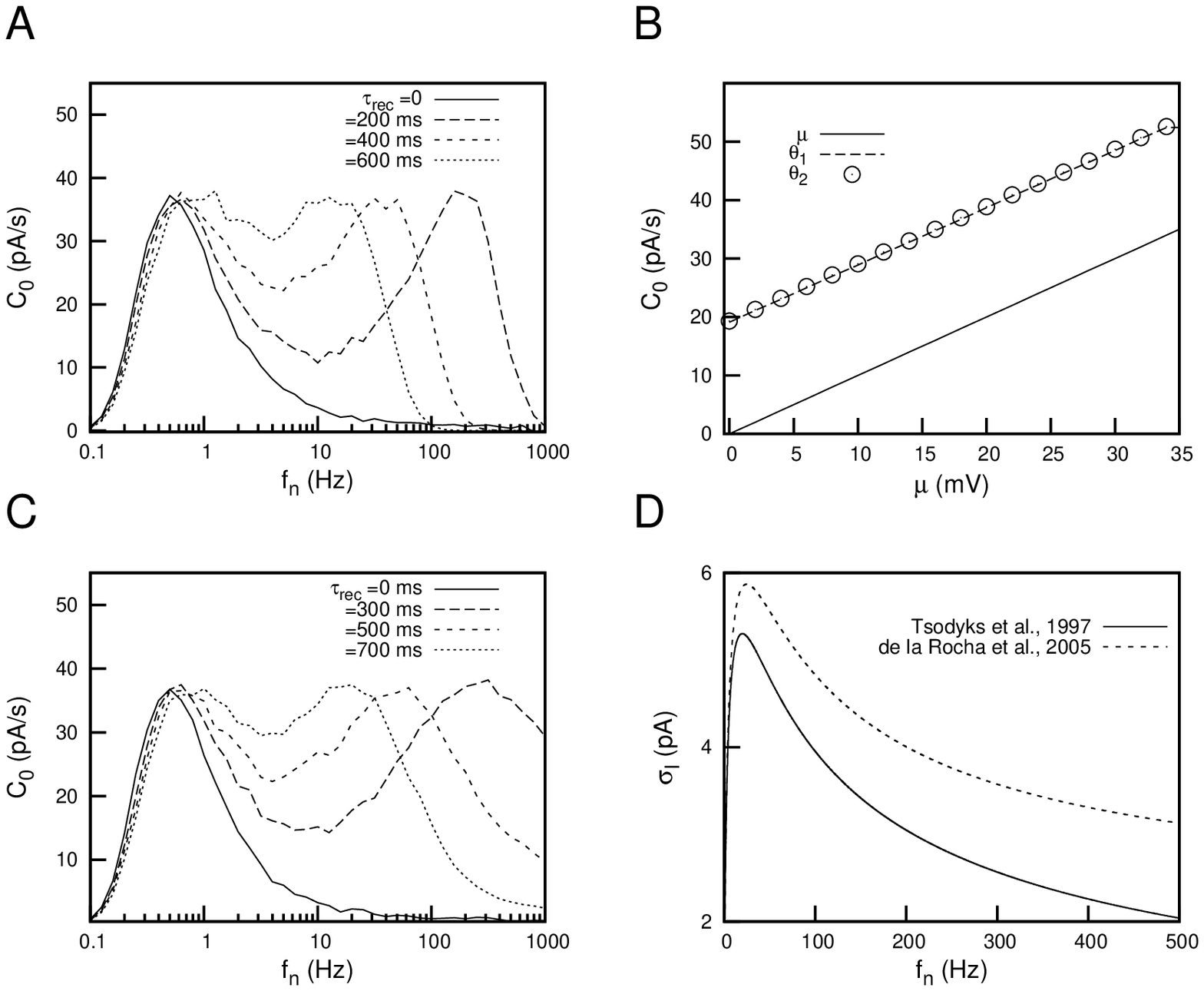,width=14cm}}
\caption{Mejias and Torres}
\label{fig6}
\end{figure}
\newpage 

\begin{figure}[t!]
\centerline{\psfig{file=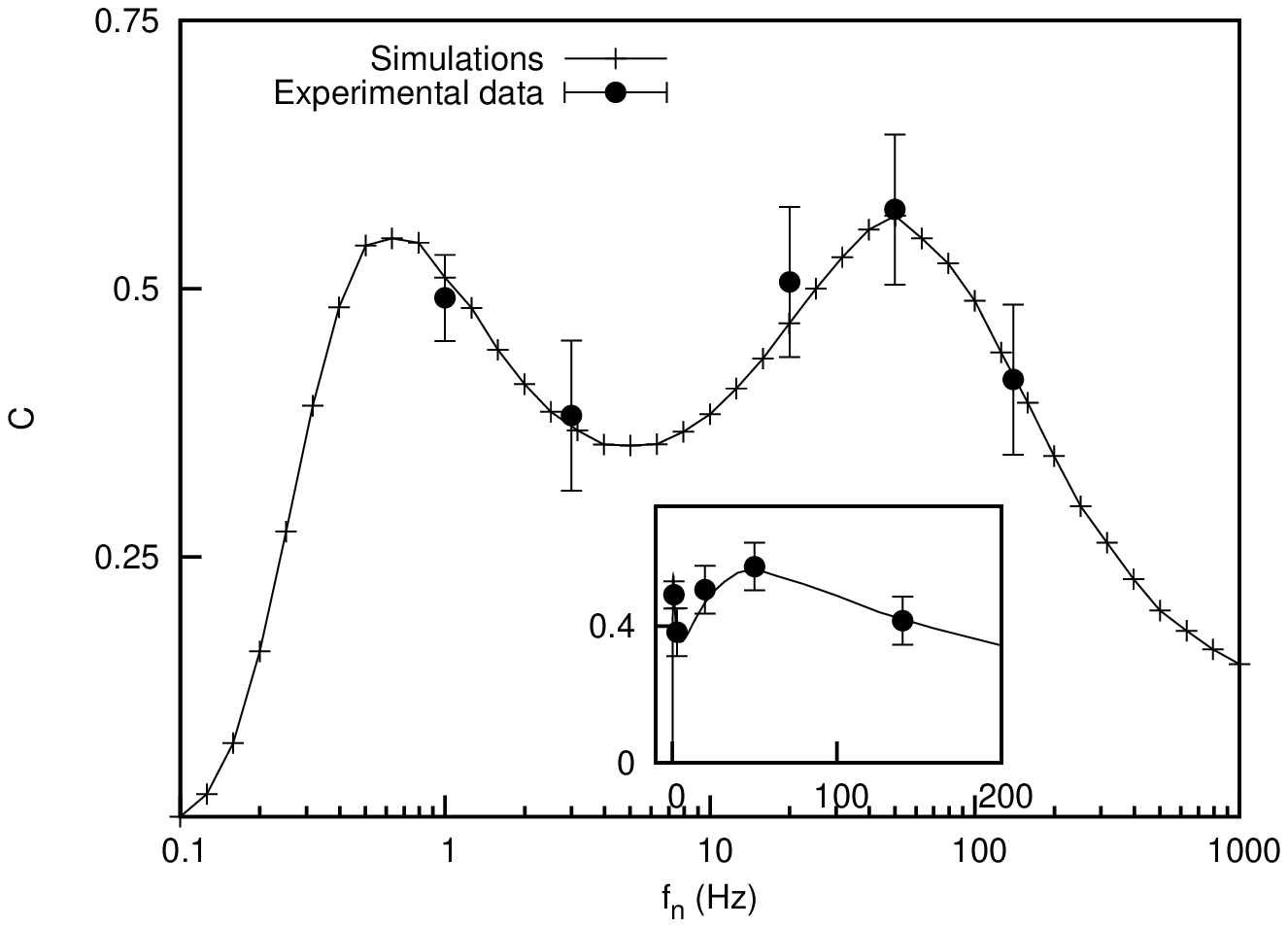,width=12cm}}
\caption{Mejias and Torres}
\label{fig7}
\end{figure}
\newpage 

\begin{figure}[t!]
\centerline{\psfig{file=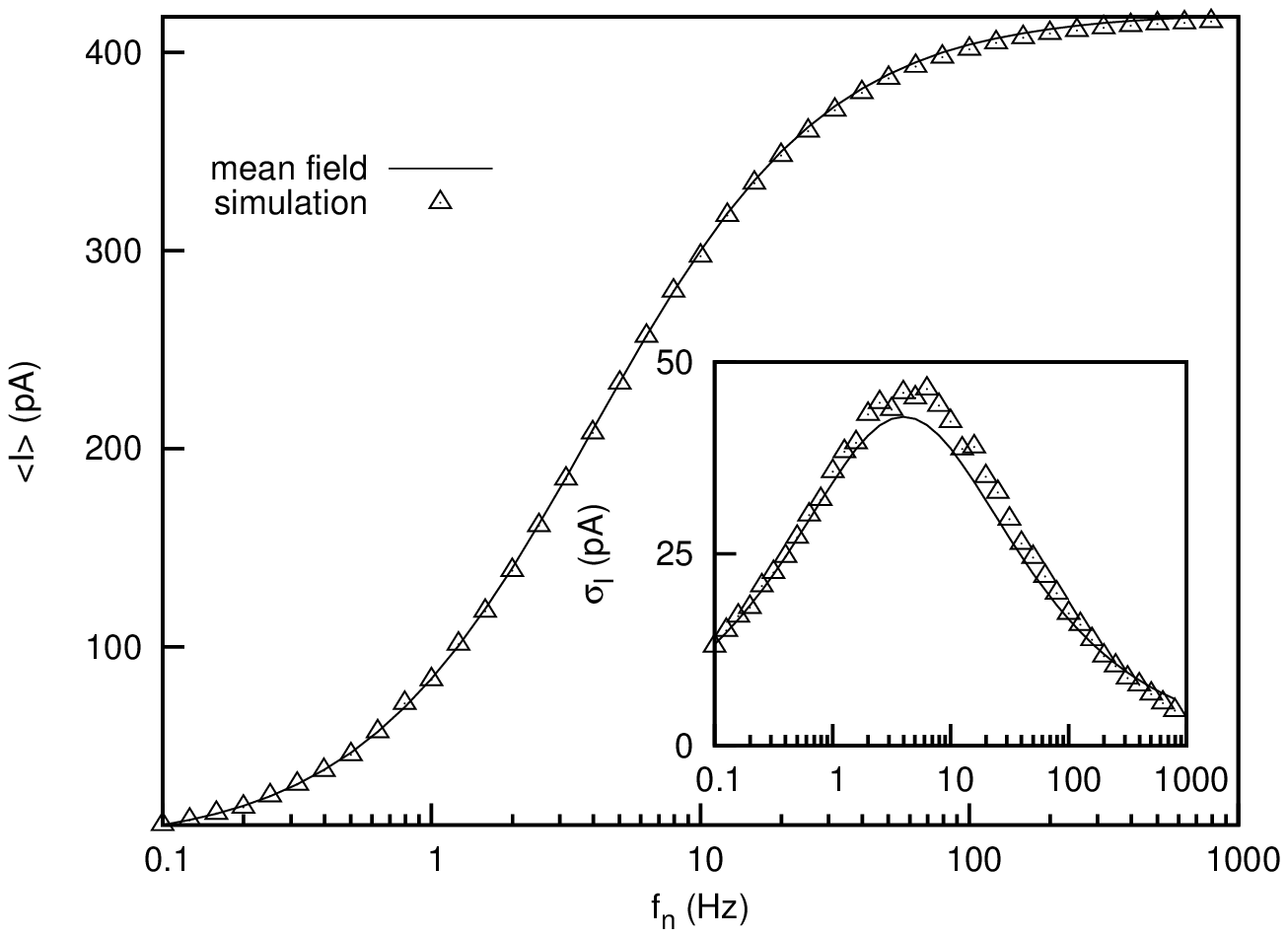,width=12cm}}
\caption{Mejias and Torres}
\label{fig8}
\end{figure}
\newpage


\end{document}